\documentstyle[12pt,sw20lart]{article}

\renewcommand{\theequation}{\thesection.\arabic{equation}}

\input{tcilatex}
\input tcilatex

\begin{document}

\titlepage
\vspace{22mm} \centerline
{\bf {\Large \ Higher Loop String Cosmology}} \centerline
{\bf {\Large \ with Moduli and}} \centerline
{\bf {\Large \ Antisymmetric Tensor Field}}

\begin{center}
\vspace{10mm} \centerline{Aram A. Saharian} \bigskip \centerline{\it %
Department of Physics, Yerevan State University} \centerline{\it 1 Alex
Manoogian St., 375049 Yerevan, Armenia} \vskip 2 cm

{\large Abstract}
\end{center}

\noindent
The classical evolution of a homogeneous cosmological model is investigated
within the framework of low - energy effective string gravity with higher
genus corrections. Various conformal frames are considered. For the general
case of correction functions in the Lagrangian we give the cosmological
solutions with arbitrary curvature and dilaton, modulus and Kalb-Ramond
fields. They generalize previously known tree level solutions. The features
of the solutions are discussed.

\bigskip

PACS Number(s): 98.80.Bp, 98.80.Cq, 98.80.Hw

\vspace{5mm}

\vfill
\newpage \renewcommand{\theequation}{1.\arabic{equation}} %
\setcounter{equation}{0}

\section{ Introduction}

Recently there has been a dramatic new development in the framework of
superstring and supersymmetric Yang-Mills theories, which leads to the
remarkable progress in understanding the non-perturbative aspects of these
theories ( for recent reviews, see \cite{schwarz}-\cite{lerche} ). Various
duality symmetries play an essential role to compute strong coupling phases.
One of the main consequences of string dualities is that the five apparently
distinct superstring theories are actually related by duality
transformations. In particular the strong coupling physics of certain
superstring theories may be reformulated as the weak coupling physics of
dual string theories (S - duality). Similar property, known as T - duality,
takes place with respect to length scales. One interesting example is the
duality in six dimensions between heterotic strings compactified on 4-torus
and type IIA superstring compactified to K3. Moreover, recent results
indicate that different types of string models correspond to different
perturbative expansions of the same underlying theory, the so - called M-
theory \cite{witten0} , whose low-energy limit is eleven dimensional
supergravity.

However, since beyond the first quantized framework of the Polyakov path
integral, our knowledge of string theory is sorely lacking, it is very
complicated to get definite results for quantum gravitational effects in
framework of these theory. Therefore the investigation of stringy effects on
the basis of low-energy effective theory becomes important. The presence of
additional degrees of freedom (dilaton, axion etc.) essentially differs such
a theory from General Relativity (GR). In particular, within the framework
of such an approach a number of interesting results are obtained in black
hole physics (see, for example \cite{banks}-\cite{kanti} and references
therein). Another area of investigations of stringy effects is the early
cosmology. The cosmological consequences of string theory and their possible
manifestations are important both for observational verification of theory
and for the resolution of a number of problems in modern cosmology.
Increasingly growing number of works in this direction (see \cite{binetruy}-%
\cite{turner} and references therein) has resulted to the formation of a new
area of cosmological investigations - String Cosmology. Recently, various
cosmological models with dilaton and antisymmetric tensor field have been
investigated to leading order in string tension as well as with
higher-curvature corrections. In most cases the universe evolves toward
strong coupling region where quantum loop corrections become important and
the tree-level effective action can not reliably describe the dynamics. This
motivates the consideration of models with loop corrections. Another
motivation is related to the fact, that the string loop effects may
naturally generate different nonmonotonic coupling functions in the
effective action and consequently a relaxation mechanism (Damour - Polyakov
mechanism) \cite{damour} by which various moduli fields are attracted
towards their present vacuum expectation values. In particular, in \cite
{damour1} it was shown that an inflationary era within Damour - Polyakov
mechanism could solve the Polonyi moduli problem.

Recently, an interesting alternative to the standard inflationary universe,
motivated by the scale factor duality \cite{venez1, tseytlin2} of the string
effective action has been developed in \cite{venez1},\cite{gasperini3} and
is known as pre-big bang inflation. According to this scenario the universe
starts from string perturbative vacuum when the string dilaton is deep in
the weak coupling region and curvature is small. Because of the instability
of this vacuum the universe inflates to a stage with strong couplings and
high curvature. It is assumed that after such an inflationary phase driven
by dilaton kinetic energy, a branch change occurs to the standard
radiation-dominated era (about the possibilities of such an exit see \cite
{brust1}-\cite{sah5}). The necessary conditions for successful exit are
related to the violations of energy conditions appearing in Hawking -
Penrose singularity theorems of GR \cite{madden1},\cite{sah5}. Loop
corrections together with higher curvature terms are natural sources
satisfying this requirement.

In the present paper we investigate multidimensional cosmological models
with variable dilaton in the framework of low-energy effective string
gravity with higher genus corrections. As an additional source a set of
Kalb-Ramond and modulus fields is considered. In section 2 the structure of
corresponding effective action and various conformal frames are discussed.
The most important frames among these ones are string Einstein and Jordan
frames. In section 3 the corresponding field equations are derived in
general conformal frame and $D$ -dimensional homogeneous anisotropic
cosmological model is considered. The equations of model take the simplest
form in Einstein frame (section 4). In particular, in this frame it is easy
to find vacuum solutions with Ricci - flat subspaces. Further, the
anisotropic solutions are considered with a set of zero potential scalar
fields as a nongravitational source. At $D$ = 4 the Kalb - Ramond field can
be reduced to scalar field (axion) as well. However for the axion, unlike
the case of usual scalar fields, at tree level the Einstein and Jordan
frames do not coincide. The set of scalar fields and Kalb - Ramond field is
equivalent to an additional source with the extreme hard equation of state.
For such a source the equations describing the evolution of cosmological
scale factors in Einstein frame decouple from the scalar field equation.
This allows to find the general solution for anisotropic model with Ricci -
flat subspaces. In section 5 the isotropic model with curved space is
considered. As in the previous case the extreme hard equation of state
allows us to find the general solution for the set of scalar fields and Kalb
- Ramond field. It generalizes previously known tree level solutions.

\renewcommand{\theequation}{2.\arabic{equation}} \setcounter{equation}{0}

\section{ String effective action and conformal frames}

Perturbative string theory contains two parameters: the string tension $%
\alpha ^{^{\prime }}$ of inverse mass squared dimension ( $h=c=1$ ) and
dimensionless string coupling constant $g_s$. The first of them determines
the length scale in the theory and controls stringy effects: at $\alpha
^{^{\prime }}\rightarrow 0$ the theory becomes the effective field theory.
The second parameter $g_s$ controls the quantum corrections and is the
parameter of loop expansion. At lowest order in string tension the effective
action can be written down as \cite{green}-\cite{kiritsis}, \cite{damour} 
\begin{equation}
S=\int d^Dx\sqrt{\left| \widetilde{G}\right| }\left[ -\widetilde{F}_R\left(
\varphi \right) \widetilde{R}-4\widetilde{F}_\varphi \left( \varphi \right)
\partial _M\varphi \widetilde{\partial }^M\varphi +\widetilde{L}_m\left(
\varphi ,\widetilde{G}_{MN},\psi \right) \right] ,  \label{actionst}
\end{equation}
where $\varphi $ is the dilaton field, $\widetilde{R}$ denotes curvature
scalar of $D$ - dimensional metric $\widetilde{G}_{MN}$, $\widetilde{L}_m$ -
is the Lagrangian density of other fields, collectively noted by a symbol $%
\psi $ and, in general, depending on the metric and dilaton (see below).
Hereafter the symbol $\sim $ above the letters specifies the quantities in
string conformal frame, the metric of which is the metric of corresponding $%
\sigma $- model. Note that the string coupling is given by the vacuum
expectation value of the dilaton field: $g_s=\left\langle e^{2\varphi
}\right\rangle $ . In the action (\ref{actionst}) the most important special
cases of Lagrangian are:

1) The antisymmetric Kalb-Ramond field $B_{MN}$ with strength $%
H_{MNP}=3\partial _{[P}B_{MN]}$ : 
\begin{equation}
\widetilde{L}_m=\frac 1{12}\widetilde{F}_H\left( \varphi \right) \widetilde{H%
}^2,\quad \widetilde{H}^2=H_{MNP}\widetilde{H}^{MNP}  \label{kalb}
\end{equation}

2) Gauge field with strength $F_{iMN}^a$: 
\begin{equation}
\widetilde{L}_m=\sum_{i,a}\frac 1{4g_i^2}\widetilde{F}_F\left( \varphi
\right) F_{iMN}^a\widetilde{F}_i^{aMN},  \label{gauge}
\end{equation}
where the index $i$ labels the various simple components of the gauge group,
and index $a$ spans the corresponding adjoint representations.

3) Dilaton potential 
\begin{equation}
\widetilde{L}_m=-\widetilde{V}\left( \varphi \right)  \label{dilpot}
\end{equation}

4) Modulus fields $\psi _i$ , which are string modes associated with
compactification of extra dimensions when $D<D_{cr}$ : 
\begin{equation}
\widetilde{L}_m=\sum_i\widetilde{F}_{\psi _i}\left( \varphi \right) 
\widetilde{G}^{MN}\partial _M\psi _i\partial _N\psi _i-U\left( \varphi ,\psi
_i\right)  \label{modulus}
\end{equation}
In the simplest cases moduli correspond to the radii of compact dimensions.

The potential terms in (\ref{dilpot}) and (\ref{modulus}) are expected to be
nonperturbative, related to the supersymmetry breaking in the theory. The
functions $\widetilde{F}_K\left( \varphi \right) ,K=R,\varphi ,H,F,\psi _i$
receive string perturbative as well as nonperturbative corrections. For the
heterotic string on perturbative level we can write the following expansion 
\begin{equation}
\widetilde{F}_K\left( \varphi \right) =e^{-2\varphi }\left(
1+\sum_{l=1}^\infty \widetilde{Z}_K^{\left( l\right) }e^{2l\varphi }\right)
\label{pertexp}
\end{equation}
where dimensionless coefficient $\widetilde{Z}_K^{\left( l\right) }$ -
represents $l$ - loop contribution, and the parameter of loop expansion is $%
e^{2\varphi }$: each loop of the string diagrams gives the contribution $%
\sim e^{2\varphi }$. At $\varphi \ll -1$ the system is in the weak coupling
region and is described by tree approximation of the string diagrams. Note
that for the type-I and type-II superstrings the kinetic terms of the R-R
fields do not have dilaton dependence at tree level.

As already it was noted above, the action (\ref{actionst}) is written in
string conformal frame. However, depending on the choice of relevant
measuring units, physical are various conformal frames (about the physical
equivalence of various conformal frames see, for example, \cite{campbell,
casas, venez2, sah1}). For example, if the act of measurement has an
electromagnetic character, in this process the metric of conformal frame is
measured, in which the electromagnetic part of the action is independent on
dilaton field (Jordan frame for electromagnetic field). From the point of
view of comparison of a dynamical picture of evolution of specific model in
various conformal frames it is convenient to consider general conformal
frame, connected to string frame by transformation of $D$ - dimensional
metric: 
\begin{equation}
G^{MN}=\Omega ^2\left( \varphi \right) \widetilde{G}^{MN}  \label{transform}
\end{equation}
with arbitrary rather smooth function $\Omega (\varphi )$. Up to divergent
terms in new conformal frame the action (\ref{actionst}) takes the form 
\begin{equation}
S=\int d^Dx\sqrt{\left| G\right| }\left[ -F_R\left( \varphi \right)
R-4F_\varphi \left( \varphi \right) \partial _M\varphi \partial ^M\varphi
+L_m\left( \varphi ,G_{MN},\psi \right) \right] ,  \label{actiongen}
\end{equation}
where the following new functions are introduced 
\begin{eqnarray}
4F_\varphi \left( \varphi \right) &=&\Omega ^{n-1}\left[ n(n-1)\left( \Omega
^{^{\prime }}/\Omega \right) ^2\widetilde{F}_R+2n\left( \Omega ^{^{\prime
}}/\Omega \right) \widetilde{F}_R^{^{\prime }}+4\widetilde{F}_\varphi \right]
\label{not1} \\
F_R\left( \varphi \right) &=&\Omega ^{n-1}\widetilde{F}_R,\quad L_m\left(
\varphi ,G_{MN},\psi \right) =\Omega ^D\widetilde{L}_m\left( \varphi ,\Omega
^2G_{MN},\psi \right) ,  \nonumber
\end{eqnarray}
$n=D-1$ - is the number of spatial dimensions, and the prime denotes
differentiation with respect to $\varphi $. The action (\ref{actiongen})
represents a generalized $D$ - dimensional scalar - tensor theory with
nongravitational Lagrangian which depends on the scalar field.

At tree level it is convenient to choose the conformal multiplier in (\ref
{transform}) as 
\begin{equation}
\Omega ^2\left( \varphi \right) =e^{c\varphi }  \label{treemult}
\end{equation}
with arbitrary constant $c$. The gravitational part of the corresponding
action is now equivalent to $D$ - dimensional Jordan - Brans - Dicke theory
with parameter \cite{casas}, \cite{sah1, sah2} 
\begin{equation}
\omega =-1+\frac{8-(n-1)c}{\left[ 4-(n-1)c\right] ^2}c  \label{omegatree}
\end{equation}
In particular, in string frame $(c=0)$ we receive from here $\omega =-1$.

The one of the most important special cases of (\ref{actiongen}) is the
action in Einstein (E-) frame, when the term with Ricci scalar has the
standard canonical form. The following choice of the conformal factor
corresponds to this frame 
\begin{equation}
\Omega ^2\left( \varphi \right) =\widetilde{F}_R^{-2/(n-1)}  \label{efactor}
\end{equation}
and the corresponding action takes the form 
\begin{equation}
S=\int d^Dx\sqrt{\left| \overline{G}\right| }\left[ -\overline{R}-4\overline{%
F}_\varphi \left( \varphi \right) \partial _M\varphi \overline{\partial }%
^M\varphi +\overline{L}_m\left( \varphi ,\overline{G}_{MN},\psi \right)
\right] ,  \label{actione}
\end{equation}
where overbars indicate that he corresponding quantities are taken in the
E-frame. Here we assumed, that $\widetilde{F}_R>0$. If this function is
negative valued, the conformal factor can be chosen similarly (\ref{efactor}%
) with the absolute value. Now in the corresponding action the Ricci scalar
will enter with positive sign, and the corresponding gravitational constant
will be negative. The function at the kinetic term of dilaton field is
connected to the corresponding function of string frame by relation 
\begin{equation}
4\overline{F}_\varphi \left( \varphi \right) =-\frac n{n-1}\left( \frac{%
\widetilde{F}_R^{^{\prime }}}{\widetilde{F}_R}\right) ^2+4\frac{\widetilde{F}%
_\varphi }{\widetilde{F}_R}.  \label{effie}
\end{equation}
For $\overline{F}_\varphi <0$ by introducing a new scalar field $\phi $ as 
\begin{equation}
d\phi =2\sqrt{-\overline{F}_\varphi }d\varphi  \label{newphi}
\end{equation}
the kinetic term can be written in canonical form. Below we shall consider
just this case. At tree level $\overline{F}_\varphi =-1/(n-1)$ and the new
scalar field is proportional to the dilaton: 
\begin{equation}
\phi =\frac 2{\sqrt{n-1}}\varphi  \label{newphi1}
\end{equation}

Another important conformal frame of scalar - tensor theories is the
so-called Jordan frame, in which the nongravitational part of the action
does not contain the scalar field. In this frame the laws of evolution of
the nongravitational fields take the same form as in multidimensional GR.
For the action (\ref{actiongen}) the Jordan frame, in general, can not be
realized. Let us consider the important special case, when in (\ref{actionst}%
) the dilaton dependence is factorized: 
\begin{equation}
\widetilde{L}_m\left( \varphi ,\widetilde{G}_{MN},\psi \right) =\widetilde{F}%
_L\left( \varphi \right) \widetilde{L}\left( \widetilde{G}_{MN},\psi \right)
.  \label{lag1}
\end{equation}
For these Lagrangians the Jordan frame exists, if the function $\widetilde{L}
$ has a certain conformal weight $\beta $: 
\begin{equation}
\widetilde{L}\left( \Omega ^2G_{MN},\psi \right) =\Omega ^{2\beta }%
\widetilde{L}\left( G_{MN},\psi \right)  \label{lag2}
\end{equation}
In general conformal frame the corresponding function is determined from the
last relation of (\ref{not1}) and has the form 
\begin{equation}
L_m=\Omega ^{D+2\beta }\widetilde{F}_L\left( \varphi \right) \widetilde{L}%
\left( G_{MN},\psi \right)  \label{lag3}
\end{equation}
Now the choice of the conformal factor in accordance with 
\begin{equation}
\Omega ^{D+2\beta }=\left| \widetilde{F}_L\left( \varphi \right) \right|
^{-1}  \label{jordan1}
\end{equation}
leads to the Jordan frame with exception of the case $\beta =-D/2$ , when $%
\widetilde{L}_m$ is conformal invariant. For examples considered above the
parameter $\beta $ has the following values: $\beta =-3$ for Kalb - Ramond
field, $\beta =-2$ for gauge field, $\beta =0$ for dilaton potential. The
Lagrangian (\ref{modulus}) has a certain conformal weight ($\beta =-1$) if
the potential terms are absent and the condition (\ref{lag1}) is fulfilled
if the functions at kinetic terms are universal. Note that for the R-R
fields the string and Jordan frames coincide at tree level.

The next conformal frame, useful for calculations of quantum gravity effects
in background field method (see for example \cite{shapiro}) is the frame
with no dilaton kinetic term in the action, corresponding to $F_\varphi
\left( \varphi \right) =0$ in (\ref{actiongen}). As it follows from relation
(\ref{not1}) such a frame exists when $\overline{F}_\varphi \leq 0$.

\renewcommand{\theequation}{3.\arabic{equation}} \setcounter{equation}{0}

\section{ Field equations and cosmological model}

By introducing a new scalar field 
\begin{equation}
\Phi =F_R\left( \varphi \right)  \label{newphi2}
\end{equation}
the action (\ref{actiongen}) can be written as 
\begin{equation}
S=\int d^Dx\sqrt{\left| G\right| }\left[ -\Phi R+\omega \left( \Phi \right)
\partial _M\Phi \partial ^M\Phi /\Phi +L\left( \Phi ,G_{MN},\psi \right)
\right] ,  \label{actionnew}
\end{equation}
where the following notation is introduced 
\begin{equation}
\omega \left( \Phi \right) =-4F_\varphi \frac{F_R}{F_R^{^{\prime }2}},\quad
L\left( \Phi ,G_{MN},\psi \right) =L_m\left( \varphi \left( \Phi \right)
,G_{MN},\psi \right)  \label{not2}
\end{equation}
The theory defined by (\ref{actionnew}) is a generalized $D$ - dimensional
scalar - tensor theory of gravity with nontrivial coupling of gravitational
scalar to the nongravitational sector. The variation of this action leads to
the following equations of motion for $G_{MN}$ and $\Phi $ (see also \cite
{will} for corresponding equations in scalar-tensor theories and \cite
{casas, sah1, sah2} for corresponding tree level equations): 
\begin{eqnarray}
R_{MN}-\frac 12G_{MN}R &=&\frac 1{2\Phi }T_{MN}+\frac 1\Phi \left(
D_MD_N\Phi -G_{MN}\Box \Phi \right) +  \label{fieldeq} \\
&&\ +\frac \omega {\Phi ^2}\left( \partial _M\Phi \partial _N\Phi -\frac
12G_{MN}\partial _P\Phi \partial ^P\Phi \right) ,  \nonumber \\
\frac{2\omega }\Phi \Box \Phi +\partial _P\Phi \partial ^P\left( \frac
\omega \Phi \right) &=&-R+\frac 1{\sqrt{\left| G\right| }}\frac{\delta
\left( L\sqrt{\left| G\right| }\right) }{\delta \Phi },  \nonumber
\end{eqnarray}
where $D_M$ denotes the covariant derivative by metric $G_{MN}$, $\Box
=G^{MN}D_MD_N$ is the covariant Dalambertian, 
\begin{equation}
T_{MN}=\frac 2{\sqrt{\left| G\right| }}\frac{\delta \left( L\sqrt{\left|
G\right| }\right) }{\delta G^{MN}}  \label{emt}
\end{equation}
is the energy - momentum tensor. Eliminating $R$ with the help of
convolution of the first equation (\ref{fieldeq}) these equations can be
written down also as 
\begin{eqnarray}
R_{MN}-\frac 1{2\Phi }\left( T_{MN}-\frac{G_{MN}}{n-1}T\right) &=&\Phi
^{-1}\left( D_MD_N\Phi +\frac{G_{MN}}{n-1}\Box \Phi \right)  \label{fieldeq2}
\\
&&\ +\frac \omega {\Phi ^2}\partial _M\Phi \partial _N\Phi , \\
2\left( \omega +\frac n{n-1}\right) \Box \Phi +\partial ^P\Phi \partial
_P\omega &=&\frac 1{n-1}\left( T+\frac{n-1}{\sqrt{\left| G\right| }}\Phi 
\frac{\delta L\sqrt{\left| G\right| }}{\delta \Phi }\right) .  \nonumber
\end{eqnarray}

As a consequence of dilaton dependence of Lagrangian $L$ the corresponding
energy-momentum tensor is acted upon by a dilaton gradient force: 
\begin{equation}
D_MT_N^M=-\frac 1{\sqrt{\left| G\right| }}\frac{\delta L\sqrt{\left|
G\right| }}{\delta \Phi }\partial _N\Phi .  \label{conteq}
\end{equation}

We shall consider $D$ - dimensional homogeneous cosmological model with
spacetime structure $R\otimes M^1\otimes ...\otimes M^P$ and with the metric 
\begin{equation}
G_{MN}=diag\left( N^2\left( t\right) ,...,-R_i^2\left( t\right)
g_{lm}^{(i)},...\right) ,  \label{cosmetric}
\end{equation}
where $M_i$ is a maximally symmetric $n_i$ dimensional space , $\sum n_i=n$, 
$g_{lm}^{(i)}$ is the metric in this space, $R_i(t)$ and $N$ are
corresponding scale factors and lapse function. Functions $N(t)$ and $R_i(t)$
,. are depending on conformal frame and are related to the corresponding
quantities in string and E - frames through the transformations 
\begin{eqnarray}
N(t) &=&\Omega ^{-1}\left( \varphi \right) \widetilde{N}(t)=\overline{\Omega 
}^{-1}\overline{N}(t),  \label{connect} \\
R_i(t) &=&\Omega ^{-1}\left( \varphi \right) \widetilde{R}_i(t)=\overline{%
\Omega }^{-1}\overline{R}_i(t),\quad \overline{\Omega }\equiv \Omega 
\widetilde{F}_R^{1/(n-1)}.
\end{eqnarray}
They relate the pictures of cosmological evolution in various frames. From
the field equations it follows, that for the metric (\ref{cosmetric}) the
energy - momentum tensor is diagonal and can be represented as 
\begin{equation}
T_N^M=diag\left( \varepsilon ,...,-\delta _n^mp_i,...\right) ,
\label{cosemt}
\end{equation}
where $\varepsilon $ is the energy density, $p_i$ is the effective pressure
in the subspace $M^i$. If the Lagrangian does not depend on derivatives of
the metric tensor, the values of these quantities in various conformal
frames are related by 
\begin{equation}
T_N^M=\Omega ^D\widetilde{T}_N^M=\overline{\Omega }^D\overline{T}_{N.}^M
\label{temen}
\end{equation}
Introducing the notation 
\begin{equation}
a_i=p_i/\varepsilon ,\quad \alpha =\frac 1{\varepsilon \sqrt{\left| G\right| 
}}\frac{\delta L\sqrt{\left| G\right| }}{\delta \varphi },\quad \overline{a}%
=1-\sum_{i=1}^pn_ia_i  \label{notai}
\end{equation}
The equations of cosmological model can be written down as 
\begin{eqnarray}
\stackrel{}{\dot{H}_i+yH_i-\delta _i\frac{b^2}2\frac{\stackrel{.}{\Phi }}%
\Phi \omega }+N^2k_i\frac{n_i-1}{R_i^2} &=&\frac{N^2}\Phi b_i\varepsilon
,\quad i=0,1,...,p  \label{cosequation} \\
\sum_{i,l=0}^pa_{il}H_iH_l+\sum k_in_iN^2\frac{n_i-1}{R_i^2} &=&\frac{N^2}%
\Phi \varepsilon  \nonumber
\end{eqnarray}
where the overdots denote time derivatives , $k_i=-1,0,1$ for subspaces with
negative, zero and positive curvatures, correspondingly, 
\begin{eqnarray}
H_i &=&\dot{R}_i/R_i,\quad i=0,1,...,p,\quad R_0\equiv \Phi ,\quad n_0=1
\label{notha} \\
y &=&\sum_{i=0}^pn_iH_i-\stackrel{.}{N}/N,\quad b^2=\left[ \omega
(n-1)+n\right] ^{-1}=-\left[ 4(n-1)\overline{F}_\varphi \right] ^{-1}\left(
F_R^{^{\prime }}/F_R\right) ^2  \nonumber \\
b_0 &=&\frac 12b^2\left[ \overline{a}+(n-1)\alpha \frac{F_R}{F_R^{^{\prime }}%
}\right] ,\quad b_i=\frac 12\left( a_i+\frac{\overline{a}}{n-1}\right) -%
\frac{b_0}{n-1},\quad i=1,...,p  \nonumber \\
\delta _0 &=&1-n,\quad \delta _1=1,\quad a_{il}=n_in_l-n_l\delta _{il},\quad
i=1,...,p,\quad a_{00}=-\omega .  \nonumber
\end{eqnarray}
At tree level and for the conformal factor (\ref{treemult}) the function $%
b\left( \varphi \right) $ is constant 
\begin{equation}
b=\frac c4(n-1)-1,\quad F_R=e^{2b\varphi }.  \label{treebe}
\end{equation}
The choice $N=1$ corresponds to the synchronous reference system in general
conformal frame, and synchronous time coordinate is depend on particular
frame. In the cosmological context the conservation equation (\ref{conteq})
takes the form 
\begin{equation}
\stackrel{.}{\varepsilon }/\varepsilon +\sum_{i=1}^pn_i(1+a_i)H_i+\alpha 
\stackrel{.}{\varphi }=0,  \label{cosconteq}
\end{equation}
where the quantities $a_i$ and $\alpha $, in general, are functions of time.
Note that in the Jordan frame $\alpha =0$. From (\ref{temen}) it follows,
that $a_i$ are conformal invariant, and the quantities $\alpha $ in various
conformal frames are connected by relations 
\begin{equation}
\alpha =\widetilde{\alpha }-\overline{a}\Omega ^{^{\prime }}/\Omega ,
\label{alfa1}
\end{equation}
if the Lagrangian $\widetilde{L}_m$ does not depend on derivatives of the
metric.

In the case of the equation of state with constant $a_i$ and for the
function $\alpha $, depending only on dilaton field , the integration of
equation (\ref{cosconteq}) yields 
\begin{equation}
\varepsilon =const\cdot \exp \left( -\int \alpha \left( \varphi \right)
d\varphi \right) \prod_{i=1}^pR_i^{-n_i(1+a_i)}  \label{epscos}
\end{equation}
In particular, for dust matter $a_i=0$ and we obtain $\varepsilon \sim 1/V$,
where $V$ is the volume of multidimensional space.

As an example of an additional source we shall consider system of scalar
fields $\psi _i$ with Lagrangian density (\ref{modulus}) for the case of
zero potential. Within the framework of homogeneous cosmological models
assuming that $\psi _i=\psi _i(t)$ we obtain for corresponding energy
density and pressure 
\begin{equation}
\varepsilon =p_i=L_m=\sum_iF_{\psi _i}\stackrel{.}{\psi }_i^2/N^2
\label{scalareps}
\end{equation}
and therefore 
\begin{equation}
a_i=1,\quad i=1,...,p,  \label{aipe}
\end{equation}
that is the system is described by the extreme hard equation of state. The
corresponding coefficients $b_i$ are determined from (\ref{notha}) and are
equal 
\begin{equation}
b_0=\frac{F_R^{^{\prime }}}{8F_R\overline{F}_\varphi }\left( \frac{%
F_R^{^{\prime }}}{F_R}-\alpha \right) ,\quad b_i=\frac{b_0}{1-n},i=1,...,p
\label{be0bei}
\end{equation}
From the equation of motion for the field $\psi _i$ 
\begin{equation}
\partial _M\left( \sqrt{\left| G\right| }F_{\psi _i}G^{MN}\partial _N\psi
_i\right) =0,  \label{psieq}
\end{equation}
we now obtain 
\begin{equation}
\stackrel{.}{\psi }_i=\frac{C_iN}{VF_{\psi _i}},\quad
V=\prod_{i=1}^pR_i^{n_i}  \label{psidot}
\end{equation}
where $C_i$ are integration constants. Hence the expressions for the energy
density and function $\alpha $ take the form 
\begin{equation}
\varepsilon =V^{-2}\sum_iC_i^2F_{\psi _i}^{-1},\quad \alpha =\frac
1{V^2\varepsilon }\sum C_i^2F_{\psi _i}^{^{\prime }}/F_{\psi _i}^2.=-\left[
\ln \left( \varepsilon V^2\right) \right] ^{^{\prime }}  \label{epsalfa}
\end{equation}
Since the function $\alpha $ depends only on dilaton field, the last two
relations can be derived also by starting directly with (\ref{epscos}).

Another important example of an additional source is Kalb - Ramond field
(see (\ref{kalb})). The corresponding equations of motion have the form 
\begin{equation}
\partial _M\left( \sqrt{\left| G\right| }F_HH^{MNP}\right) =0,\quad
F_H=\Omega ^{D-6}\widetilde{F}_H.  \label{kalbeq}
\end{equation}
In $D=4$ these equations are solved by the ansatz 
\begin{equation}
H^{MNP}=\frac 1{\sqrt{\left| G\right| }}F_H^{-1}\varepsilon ^{MNPQ}\partial
_Qh,  \label{ansatz}
\end{equation}
where $\varepsilon ^{MNPQ}$ is completely antisymmetric 4-tensor, $h$ is the
pseudoscalar field of axion. From the Bianci identity $\partial
_{[Q}H_{MNP]}=0$ the equation of motion for the field $h$ follows: 
\begin{equation}
\Box h-\partial ^Mh\partial _MF_H/F_H=0  \label{axioneq}
\end{equation}
The corresponding Lagrangian density has the form 
\begin{equation}
L_m=F_h\partial ^Mh\partial _Mh,\quad F_h=1/\left( 2F_H\right)
\label{axionlag}
\end{equation}
in general conformal frame related to the string one by transformation (\ref
{transform}). Thus in this case the Kalb - Ramond field is equivalent to the
pseudoscalar field $h$ and the corresponding formulas are a special case of
the previous example. However it is necessary to note, that if for usual
scalar fields in the string frame at tree level $\widetilde{F}_{\psi
_i}=e^{-2\varphi }$, for the heterotic string axion, as it follows from (\ref
{axionlag}), $\widetilde{F}_h=e^{2\varphi }/2$ and for the type-I axion $%
\widetilde{F}_h=1/2$.

\renewcommand{\theequation}{4.\arabic{equation}} \setcounter{equation}{0}

\section{ Solutions with Ricci - flat subspaces}

The cosmological field equations are simplest in the Einstein metric. By
introducing a new scalar field $\phi $ in accordance with (\ref{newphi}),
the system of equations (\ref{cosequation}) can be written as (here and
below, to not complicate the formulas,we omit the overbars over the letters
indicating the quantities in the E - frame; everywhere, where it is not
especially noted, the E - frame is considered) 
\begin{eqnarray}
\stackrel{}{\stackrel{.}{H}_i+\left( \sum_{i=1}^pn_iH_i-\stackrel{.}{N}%
/N\right) H_i} &=&N^2b_i\varepsilon -N^2k_i\frac{n_i-1}{R_i^2},\quad
i=1,...,p  \label{eframeeq} \\
\stackrel{..}{\phi }+\left( \sum_{i=1}^pn_iH_i-\stackrel{.}{N}/N\right) 
\stackrel{.}{\phi } &=&\frac 12N^2\alpha _\phi \varepsilon  \nonumber \\
N^2\varepsilon +\stackrel{.}{\phi }^2-\sum k_in_iN^2\frac{n_i-1}{R_i^2}
&=&\sum_{i,l=1}^pa_{il}H_iH_l  \nonumber
\end{eqnarray}
where now in E - frame 
\begin{equation}
b_i=\frac 12\left( a_i+\frac{\overline{a}}{n-1}\right) ,\quad \alpha _\phi
=\frac 1{\varepsilon \sqrt{\left| G\right| }}\frac{\delta L\sqrt{\left|
G\right| }}{\delta \phi }  \label{eframebe}
\end{equation}
From the second equation of (\ref{eframeeq}) it follows, that for the
solutions with constant dilaton $\varphi =\varphi _0$, the values $\varphi
_0 $ have to be roots of the equation $\alpha \left( \varphi \right) =0$.
The dilaton can be fixed both by dilaton potential, or by nontrivial
dependence of Lagrangian of the other fields on dilaton (see, for example, (%
\ref{kalb}) and (\ref{gauge})). The second way is known as Damour - Polyakov
mechanism \cite{damour}. The solutions with constant dilaton coincide with
corresponding solutions of multidimensional GR. Let us consider the behavior
of the cosmological models in the neighborhood of these solutions for the
case of a set of zero potential scalar fields as a nongravitational source
(see (\ref{scalareps})) (on the relaxation of scalar - tensor isotropic
cosmological models to the solutions with constant scalar field see \cite
{nordvedt} and references therein). For such a source it is convenient to
introduce a new variable $\tau $ in accordance with 
\begin{equation}
d\tau =\frac NVdt  \label{tau2}
\end{equation}
By taking into account (\ref{epscos}) and (\ref{epsalfa}) we obtain from the
second equation (\ref{eframeeq}) an evolution equation for $\phi $ as
function of $\tau $: 
\begin{equation}
\frac{d^2\phi }{d\tau ^2}=-\frac 12\frac d{d\phi }\left( \varepsilon
V^2\right)  \label{newform}
\end{equation}
As it follows from here for a solution with constant scalar field $\phi
=\phi _0$ the value $\phi _0$ has to be an extremum of the function $%
\varepsilon V^2$: 
\begin{equation}
\frac d{d\phi }\left( \varepsilon V^2\right) =\frac d{d\phi }\sum_i\frac{%
C_i^2}{F_{\psi _i}}=0,\quad \phi =\phi _0  \label{dilconst0}
\end{equation}
The equation (\ref{newform}) can be thought of as describing a one
dimensional dynamics of a particle with a potential term $V^2\varepsilon /2$%
. The singular points of the corresponding autonomous dynamical system are
determined by equation (\ref{dilconst0}). On the phase plane $\left( \phi
,d\phi /dt\right) $ these points are saddle points for $\omega _0\equiv
\left( \frac{d^2}{d\phi ^2}V^2\varepsilon \right) _{\phi =\phi _0}<0$ and
center for $\omega _0>0$. For both these cases the dilaton dependence of the
Lagrangian for modulus and antisymmetric tensor fields can not drive the
cosmological model toward a state with constant dilaton. For the isotropic
flat cosmological model this result follows also from the analysis of Ref. 
\cite{nordvedt},\cite{damour} and is a consequence of the fact that for the
source with equation of state $\varepsilon =p$ the corresponding friction
term in the scalar field equation vanishes. Note that for the case $\omega
_0>0$ in the neighborhood of singular point the trajectories on the phase
plane $\left( \phi ,d\phi /dt\right) $ are closed curves and the
corresponding solutions have the form $\phi -\phi _0\sim \cos \left( \sqrt{%
\omega _0}\tau +\tau _0\right) $.

\subsection{Pure gravi - dilaton solutions}

We start with the simple case of Ricci - flat subspaces ($k_i=0$) and zero
nongravitational sources ($L_m=0$). From the system (\ref{eframeeq}) we find 
\begin{equation}
H=const\cdot N/V,\quad \stackrel{.}{\phi }=const\cdot N/V  \label{vachubble}
\end{equation}
Introducing E - frame synchronous time coordinate $t_E$: 
\begin{equation}
dt_E=N(t)dt  \label{detee}
\end{equation}
the solutions to cosmological equations can be written as 
\begin{equation}
H_i=\frac{H_{i0}}{t_E-t_0},\quad R_i=R_{i0}\left| t_E-t_0\right|
^{H_{i0}},\quad \sum_{i=1}^pn_iH_{i0}=1  \label{vacsol1}
\end{equation}
\begin{equation}
\phi =\phi _{10}\ln \left| t_E-t_0\right| +\phi _0,\quad \phi
_{10}^2+\sum_{i=1}^pn_iH_{i0}^2=1  \label{vacsol2}
\end{equation}
where $H_{i0},R_{i0},\phi _0,t_{0\text{ }}$- are integration constants. The
last relation is a consequence of the constraint equation. Then $\phi
_{10}=0 $ from (\ref{vacsol1}),(\ref{vacsol2}) one obtains the
multidimensional generalization of the Kasner solution of the $D$ -
dimensional GR. Note that unlike the case of Kasner solution, the solution (%
\ref{vacsol1}),(\ref{vacsol2}) has an isotropic limit when one obtains $%
H_{10}=1/n,\quad \phi _{10}=\pm \sqrt{1-1/n}$ .

For given functions $\widetilde{F}_\varphi \left( \varphi \right) $ and $%
\widetilde{F}_R\left( \varphi \right) $ the time dependence of dilaton field
is determined from 
\begin{equation}
\phi _{10}\ln \left| t_E-t_0\right| =2\int \sqrt{-\overline{F}_\varphi }%
d\varphi ,  \label{dilontime1}
\end{equation}
where the integrand can be expressed through the previous functions by the
relation (\ref{effie}). Note that the behavior of dilaton field is always
monotonic if the function $\overline{F}_\varphi $ does not change the sign.
The corresponding solution in string frame can be found by transformation 
\begin{equation}
\widetilde{R}_i=\widetilde{F}_R^{1/(1-n)}R_i,\quad dt_s=\widetilde{F}%
_R^{1/(1-n)}dt_E  \label{etostring}
\end{equation}
where $t_s$ is string frame synchronous time coordinate. At tree level the
solution of string frame takes the form \cite{mueller, venez1} 
\begin{eqnarray}
\varphi -\varphi _0 &=&\varphi _{10}\ln \left| t_s-t_{s0}\right| ,\quad 
\widetilde{R}_i=\widetilde{R}_{i0}\left| t_s-t_{s0}\right| ^{\widetilde{H}%
_{i0}}  \label{phierst} \\
\sum_{i=1}^pn_i\widetilde{H}_{i0} &=&1+2\varphi _{10},\quad \sum_{i=1}^pn_i%
\widetilde{H}_{i0}^2=1  \nonumber
\end{eqnarray}
The same solution in general conformal frame can be found in \cite{sah3}.
The corresponding isotropic solution has the form 
\begin{equation}
\varphi =\varphi _0+\frac 12\left( \pm \sqrt{n}-1\right) \ln \left|
t_s-t_{s0}\right| ,\quad H_i=\frac{\pm 1}{\sqrt{n}\left( t_s-t_{s0}\right) }
\label{isotnew1}
\end{equation}
In the case $t_s<t_{s0}$ and for the lower sign, (\ref{isotnew1}) describes
superinflationary expansion, when the evolution of the universe starts at $%
t_s=-\infty $ and dilaton is deep in weak coupling region. Such an expansion
is driven by kinetic energy of the dilaton field and corresponds to the
pre-big bang phase of the mechanism for inflationary evolution proposed in 
\cite{venez1}, \cite{gasperini3}.

\subsection{Solutions with modulus and Kalb - Ramond fields}

In the equations (\ref{eframeeq}) as an additional source we shall consider
a set of scalar fields and Kalb - Ramond field: 
\begin{equation}
L_m=\sum_iF_{\psi _i}\partial ^M\psi _i\partial _M\psi _i+\frac
1{12}F_H\left( \varphi \right) H^2  \label{lagsource}
\end{equation}
As it has been mentioned already, if we accept the ansatz (\ref{ansatz}),
the Kalb - Ramond field is also reduced to the scalar field with Lagrangian (%
\ref{axionlag}) . In particular, in cosmological context we shall assume 
\begin{equation}
\varepsilon =p_i=L_m=\sum_i^{^{\prime }}F_{\psi _i}\stackrel{.}{\psi }%
_i^2/N^2\equiv \sum_iF_{\psi _i}\stackrel{.}{\psi }_i^2/N^2+F_h\stackrel{.}{h%
}^2/N^2  \label{cossource}
\end{equation}
and prime specifies the fact that the corresponding sum includes also the
contribution of the axion field. For the functions $F_{\psi _i}$ in
accordance with (\ref{transform}) and (\ref{efactor}) we have in E - frame 
\begin{equation}
F_{\psi _i}=\widetilde{F}_{\psi _i}/\widetilde{F}_R,F_h=\left( 2\widetilde{F}%
_H\widetilde{F}_R\right) ^{-1}.  \label{efpsie}
\end{equation}
It follows from here, that at tree level of the heterotic superstring,
thanks to the universality of the functions $\widetilde{F}_k\left( \varphi
\right) $: 
\begin{equation}
F_{\psi _i}=1,\quad F_h=e^{4\varphi }/2  \label{efpsitree}
\end{equation}
For the type-I superstring we have $F_h=e^{2\varphi }/2$. Thus, for usual
scalar fields (it concerns also to the dilaton field) at tree level the E -
and Jordan frames coincide, while for axion field this is not the case. By
taking into account these relations one obtains the following asymptotic
behavior of potential term of (\ref{newform}) in weak coupling region: 
\begin{equation}
\varepsilon V^2/2=\sum_i^{^{\prime }}\frac{C_i^2}{2F_{\psi _i}}\sim \frac
12\sum_iC_i^2+C_h^2e^{-4\varphi },\quad \varphi \ll -1  \label{potasymp}
\end{equation}
and the contribution of the axion field dominates. It follows from (\ref
{cossource}), that for a source with the Lagrangian (\ref{lagsource}) all
coefficients $a_i$ are equal to 1, and therefore as we can see from (\ref
{eframebe}) all coefficients $b_i$ are equal to zero. For these values and
for Ricci - flat subspaces the solution of the first equation (\ref{eframeeq}%
) for scale factors in synchronous system of coordinates is still determined
by the relation (\ref{vacsol1}). From the constraint equation we now have 
\begin{equation}
\varepsilon +\stackrel{.}{\phi }^2=A\left( t_E-t_0\right) ^{-2},\quad
A\equiv 1-\sum_{i=1}^pn_iH_{i0}^2,  \label{eps2}
\end{equation}
(note that, if the all subspaces are expanding then the constant $A$ is
positive) where the energy density is determined by the relation (see (\ref
{epsalfa})) 
\begin{equation}
\varepsilon =\frac 1{V_0^2\left( t_E-t_0\right) ^2}\sum_i^{^{\prime }}\frac{%
C_i^2}{F_{\psi _i}}.  \label{eps3}
\end{equation}
The substitution of this expression in (\ref{eps2}) and integration of the
obtained equation leads to the following result 
\begin{equation}
\ln \left| t_E-t_0\right| =\pm \int \left[ A-\sum_i^{^{\prime
}}C_i^2V_0^{-2}/F_{\psi _i}\right] ^{-1/2}d\phi .  \label{solforphi}
\end{equation}
Here the functions $F_{\psi _i}$ are related to the corresponding functions
of the string frame by (\ref{efpsie}). Further for given function $\varphi
\left( t_E\right) $ from the equations (\ref{psidot}) one finds 
\begin{equation}
\psi _i=\int \frac{C_idt_E}{V_0F\psi _i\left( \varphi \right) \left|
t_E-t_0\right| }.  \label{depsidete}
\end{equation}
For the given coupling functions in (\ref{actionst}) the formulas (\ref
{vacsol1}), (\ref{efpsie}), (\ref{solforphi}), (\ref{depsidete}) determine
the evolution of the corresponding cosmological model in the E-frame. The
string frame solutions can be found by transformations (\ref{etostring}).
Note that unlike the case of the E-frame, the solutions for string frame
scale factors do not coincide with corresponding vacuum solutions.

As has been previously mentioned, for the solutions with constant dilaton $%
\varphi =\varphi _0$ one has $\alpha \left( \varphi _0\right) =0$. For such
a solution together with (\ref{dilconst0}) we have the following relation
between the integration constants 
\begin{equation}
\sum_i^{^{\prime }}C_i^2V_0^{-2}/F_{\psi _i}\left( \varphi _0\right) =A
\label{dilconst1}
\end{equation}
and the formula 
\begin{equation}
\psi _i=\psi _{i0}+\frac{C_i}{V_0F_{\psi _i}\left( \varphi _0\right) }\ln
\left| t_E-t_0\right|  \label{dilconst2}
\end{equation}
for modulus and axion fields. By taking into account (\ref{vacsol1}) and (%
\ref{tau2}) in the neighborhood of this solution we receive 
\begin{equation}
\phi -\phi _0\sim \cos \left( \frac{\sqrt{\omega _0}}{V_0}\ln \left|
t_E-t_0\right| \right)  \label{osc4}
\end{equation}

At tree level taking into account the expressions (\ref{efpsitree}) for
dilaton we find from (\ref{solforphi}) 
\begin{eqnarray}
e^{2\varphi } &=&\frac{A_1}2\left( \left| t_E-t_0\right| ^{\alpha _1}+\frac{%
\alpha _2}{A_1^2}\left| t_E-t_0\right| ^{-\alpha _1}\right)  \label{e2phi} \\
\alpha _1 &=&\pm \left[ \left( A-\sum_iC_i^2V_0^{-2}\right) (n-1)\right]
^{1/2},\quad \alpha _2=2C_h^2(n-1)/\alpha _1^2V_0^2,  \nonumber
\end{eqnarray}
$A_1$- is an integration constant (for the type-I superstring one has $%
e^\varphi $ in the left hand-side and $\alpha _1/2$ in the right hand side).
For nonzero antisymmetric tensor field from (\ref{e2phi}) one obtains a
restriction on possible values of dilaton field: $e^{4\varphi }\geq \alpha
_2 $. At the same approximation from the equations (\ref{depsidete}) we
obtain 
\begin{eqnarray}
\psi _i &=&\psi _{i0}+\frac{C_i}{V_0}\ln \left| t_E-t_0\right|
\label{psihash} \\
h &=&h_0-\frac{4C_h}{\alpha _1V_0A_1^2}\left[ \left| t_E-t_0\right|
^{2\alpha _1}+\alpha _2/A_1^2\right] ^{-1}  \nonumber
\end{eqnarray}
with integration constants $\psi _{i0}$ and $h_0$. The corresponding
solutions in string frame can be found by transformation (\ref{etostring}).
Here we shall consider a simple case of zero Kalb - Ramond field,
corresponding to the value of constant $\alpha _2=0$ ($C_h=0$). At tree
level the synchronous time coordinates in string and E - frames are related
by 
\[
t_s-t_{s0}=const\cdot \left| t_E-t_0\right| ^{1+\alpha _1/\left( n-1\right)
} 
\]
and the solutions of string frame are 
\begin{eqnarray}
\widetilde{R}_i &\sim &\left| t_s-t_{s0}\right| ^{\widetilde{H}_{i0}},\quad
e^{2\varphi }\sim \left| t_s-t_{s0}\right| ^{\widetilde{\alpha }_1}
\label{stringer} \\
e^{\psi _i} &\sim &\left| t_s-t_{s0}\right| ^{\psi _{i1}},\quad
\sum_{i=1}^pn_i\widetilde{H}_{i0}=1+\widetilde{\alpha }_1,\quad
\sum_{i=1}^pn_i\widetilde{H}_{i0}^2+\sum_i\psi _{i1}^2=1,  \nonumber
\end{eqnarray}
where the constants can be expressed via the previously defined constants by
relations 
\begin{eqnarray}
\widetilde{\alpha }_1 &=&\frac{\alpha _1}{1+\alpha _1/(n-1)},\quad \psi
_{i1}=\frac{\widetilde{\alpha }_1C_i}{\alpha _1V_0}  \label{newconst4} \\
\widetilde{H}_{i0} &=&\left( H_{i0}+\frac{\alpha _1}{n-1}\right) \frac{%
\widetilde{\alpha }_1}{\alpha _1}  \nonumber
\end{eqnarray}
For the pure gravi - dilaton case we have $\psi _{i1}=0$ and (\ref{stringer}%
) coincides with (\ref{phierst}).

\renewcommand{\theequation}{5.\arabic{equation}} \setcounter{equation}{0}

\section{ Isotropic models with curved space}

In the previous section we have considered multidimensional anisotropic
cosmological model with scalar and Kalb - Ramond fields for the case of
Ricci - flat subspaces. We shall now consider the isotropic models with
curved space. In E - frame the corresponding system of cosmological
equations can be obtained from (\ref{eframeeq}): 
\begin{eqnarray}
\stackrel{.}{H}+H(nH-\stackrel{.}{N}/N) &=&N^2b_1\varepsilon -N^2k\frac{n-1}{%
R^2}  \label{isoteq} \\
\stackrel{..}{\phi }+\stackrel{.}{\phi }(nH-\stackrel{.}{N}/N) &=&\frac
12N^2\alpha _\phi \varepsilon ,  \nonumber
\end{eqnarray}
where 
\begin{equation}
b_1=\frac{1-a_1}{2(n-1)},\quad a_1=\frac{p_1}\varepsilon  \label{not51}
\end{equation}
The constraint equation has the form 
\begin{equation}
N^2\varepsilon +\stackrel{.}{\phi }^2=n(n-1)\left( H^2+kN^2/R^2\right) .
\label{consteq5}
\end{equation}
The solution of cosmological equations becomes essentially simple for a
source with the extreme hard equation of state $\varepsilon =p_1$ ($a_1=1$).
In particular, as it was shown in the previous section, this condition is
fulfilled for a set of scalar fields and Kalb - Ramond field with the
Lagrangian (\ref{lagsource}). For such sources the first equation of system (%
\ref{isoteq}) coincides with the corresponding vacuum equation and is solved
most simply in terms of conformal time $\eta $, corresponding to the gauge $%
N=R$: 
\begin{equation}
ds^2=R^2\left( d\eta ^2-dl^2\right) .  \label{interval}
\end{equation}
Now by integrating the first equation of (\ref{isoteq}) for Hubble function
we find 
\begin{equation}
H^2=B/R^{2(n-1)}-k  \label{hubble5}
\end{equation}
where $B$ is an integration constant. It follows from here, that for closed
models ($k=1$) scale factor varies in finite interval $0\leq R\leq R_m$,
where $R_m=B^{1/2(n-1)}$. Making use of Eq. (\ref{hubble5}) from the
constraint equation one obtains 
\begin{equation}
N^2\varepsilon +\left( \frac{d\phi }{d\eta }\right) ^2=\frac{n(n-1)}{%
R^{2(n-1)}}B  \label{consteq51}
\end{equation}
and therefore if the energy density is nonnegative, then $B\geq 0$ . Note,
that the solutions with negative $B$ can be realized only for models with
spaces of negative curvature, and in this case the scale factor varies
within limits 
\begin{equation}
\left| B\right| ^{1/2(n-1)}\equiv R_{\min }\leq R<\infty  \label{erinterv}
\end{equation}
and the corresponding models are nonsingular: 
\begin{equation}
R=R_{\min }\left[ ch(n-1)\eta \right] ^{1/(n-1)}.  \label{negativbe}
\end{equation}
For the solutions with non-negative energy density, solving the equation (%
\ref{hubble5}) for the scale factor we have 
\begin{equation}
R=R_m\left| \frac 1{\sqrt{k}}\sin \left[ \sqrt{k}(n-1)\eta \right] \right|
^{1/(n-1)}  \label{er5}
\end{equation}

In view of this from the constraint equation we find for energy density 
\begin{equation}
N^2\varepsilon +\left( \frac{d\phi }{d\eta }\right) ^2=\frac{kn(n-1)}{\sin
^2\left[ \sqrt{k}(n-1)\eta \right] }.  \label{consteq52}
\end{equation}
In what follows we shall assume $0\leq (n-1)\eta \leq \pi $ for $k=1$ and $%
0\leq (n-1)\eta <\infty $ for $k=0,-1$. The solutions in the other regions
can be obtained fro here. Let us consider as a nongravitational source the
system with the Lagrangian density (\ref{lagsource}) and energy density (\ref
{cossource}). Below we shall not specify the value of dimension $n$ of
space. The substitution (\ref{cossource}) into the constraint equation (\ref
{consteq52}) leads to the following equation 
\begin{equation}
\frac{d\phi }{d\eta }=\frac{\pm \sqrt{k}}{\sin \left[ \sqrt{k}(n-1)\eta
\right] }\left[ n(n-1)-R_m^{2(1-n)}\sum_i^{^{\prime }}\frac{C_i^2}{F_{\psi
_i}}\right] ^{1/2}  \label{dephi5}
\end{equation}
By integrating this equation for dilaton field we find the following
relation 
\begin{eqnarray}
\pm \ln \left| \frac 1{\sqrt{k}}tg\left[ \sqrt{k}(n-1)\frac \eta 2\right]
\right| &=&(n-1)\int \left[ \frac n{n-1}\left( \frac{\widetilde{F}%
_R^{^{\prime }}}{\widetilde{F}_R}\right) ^2-\frac{4\widetilde{F}_\varphi }{%
\widetilde{F}_R}\right] ^{1/2}  \label{phifunc5} \\
&&\ \times \left[ n(n-1)-R_m^{2(1-n)}\sum_i^{^{\prime }}C_i^2\frac{%
\widetilde{F}_R}{\widetilde{F}_{\psi _i}}\right] ^{-1/2}d\varphi  \nonumber
\end{eqnarray}
For given functions $\widetilde{F}_K\left( \varphi \right) $ in the
Lagrangian (\ref{actionst}) this formula together with (\ref{er5})
determines the dynamics of corresponding cosmological model. For models with
negative energy density the evolution of dilaton field is determined from
the relation 
\begin{equation}
\pm 2arctg\left[ e^{(n-1)\eta }\right] =(n-1)\int \left[ \sum_i^{^{\prime }}%
\frac{C_i^2}{F_{\psi _i}}R_{\min }^{2(1-n)}-n(n-1)\right] ^{-1/2}d\phi
\label{phineg5}
\end{equation}
and the scale factor has the form (\ref{negativbe}). Further we shall
consider models with non- negative energy density. For given function $%
\varphi \left( \eta \right) $, determined from (\ref{phifunc5}), the
dependence of the other fields on time is determined from the equation (\ref
{psidot}): 
\begin{equation}
\frac{d\psi _i}{d\eta }=\frac{C_i}{R_m^{n-1}}\frac{\sqrt{k}F_{\psi
_i}^{-1}\left( \varphi \right) }{\sin \left[ \sqrt{k}(n-1)\eta \right] }.
\label{depsi5}
\end{equation}
In particular, for $\psi _i=h$ one obtains the equation for the axion field.
For the given coupling functions, the expressions (\ref{er5}), (\ref
{phifunc5}), (\ref{depsi5}) determine the evolution of isotropic
cosmological models with $k=0,\pm 1$ in terms of conformal time coordinate $%
\eta $.

In what follows it proves convenient to define the new variable $z$
according to 
\begin{equation}
z=\left| \frac 1{\sqrt{k}}tg\left[ \sqrt{k}(n-1)\eta /2\right] \right| .
\label{zet}
\end{equation}
The formula (\ref{er5}) for the scale factor can be now written down as 
\begin{equation}
R=R_0\left( \frac z{1+kz^2}\right) ^{1/(n-1)}.  \label{erzet}
\end{equation}
The above mentioned solutions are written through conformal time $\eta $ ,
which is connected to E - frame synchronous time coordinate by relation 
\begin{equation}
t_E-t_0=\pm \int Rd\eta =\pm \frac{2R_0}{n-1}\int_0^z\frac{z^{1/(n-1)}dz}{%
\left( 1+kz^2\right) ^{n/(n-1)}},  \label{te5}
\end{equation}
where the integral in the right hand side can be expressed through
hypergeometric function. The corresponding solutions in string frame can be
found by transformation (\ref{etostring}). For open models ($k=-1$) the
variable $z$ varies within limits $0\leq z\leq 1$, thus, as it follows from (%
\ref{te5}), time coordinate is not limited. In the case of closed models $%
0\leq z<\infty $, and $t_E$ varies in finite limits: 
\begin{equation}
\left| t_E-t_0\right| \leq t_{em}\equiv \frac{R_0}{n-1}B\left( \frac
n{2(n-1)},\frac n{2(n-1)}\right) ,  \label{liftime}
\end{equation}
where $B(x,y)$ - is Euler $\beta $-function. For the models with a flat
space from (\ref{erzet}) we have $R\sim \left| t_E-t_0\right| ^{1/n}$, which
is the isotropic limit of the solution (\ref{vacsol1}). In the limit $%
t_E\rightarrow t_0$ the contribution of curvature terms is negligible and
the behavior of open and closed models is the same as in a flat case. For an
open model and $t_E\rightarrow \infty $ from (\ref{erzet}) and (\ref{te5})
we have $R\sim t_E$.

For the solutions with constant dilaton $\varphi =\varphi _0$ the value $%
\varphi _0$ has to be a solution of the equation (\ref{dilconst0}) and from (%
\ref{dephi5}) we obtain the following relation between integration constants 
\begin{equation}
\sum_i^{^{\prime }}\frac{C_i^2}{F_{\psi _i}\left( \varphi _0\right) }%
R_m^{2(1-n)}=n(n-1)  \label{dilconst3}
\end{equation}
and the expression 
\begin{equation}
\psi _i=\frac{C_i}{(n-1)F_{\psi _i}\left( \varphi _0\right) }R_m^{1-n}\ln
z+const  \label{dilconst4}
\end{equation}
for modulus and axion fields. In the neighborhood of this solution for
models with variable dilaton one obtains $\phi -\phi _0\sim \cos \left( 
\sqrt{\omega _0}R_m^{1-n}\frac{\ln z}{n-1}\right) $.

We now turn to the consideration of the obtained solutions at tree level of
the heterotic string ( the corresponding type-I solutions can be obtained by
replacements $e^{2\varphi }\rightarrow e^\varphi ,\beta _1\rightarrow \beta
_1/2$), when the corresponding E - frame functions are defined by relations (%
\ref{efpsitree}). The integration of (\ref{depsi5}) now leads to the result 
\begin{equation}
\psi _i=\psi _{i1}\ln z+\psi _{i0},\quad \psi _{i1}=\frac{C_i}{(n-1)R_m^{n-1}%
}.  \label{psitree5}
\end{equation}
Similarly, from the relation (\ref{phifunc5}) for dilaton field we find 
\begin{equation}
e^{2\varphi }=\left( A_1z^{\beta _1}+\beta _2z^{-\beta _1}/A_1\right) /2,
\label{e2phi5}
\end{equation}
where $A_1$ is an integration constant, and $\beta _i$ are connected to the
earlier introduced constants by relations 
\begin{eqnarray}
\beta _1 &=&\pm \left[ n-R_m^{2(1-n)}\sum_iC_i^2/(n-1)\right] ^{1/2},
\label{beta12} \\
\beta _2 &=&2C_h^2/\left[ n(n-1)R_m^{2(n-1)}-\sum_iC_i^2\right] .  \nonumber
\end{eqnarray}
As it follows from (\ref{e2phi5}) for models with nonzero Kalb - Ramond
field the values of dilaton are limited by inequality $e^{4\varphi }\geq
\beta _2$. From here as a necessary condition of existence of weak coupling
region we obtain: $\beta _2\ll 1$. By taking into account the tree level
expression $F_h=e^{4\varphi }/2$ from the equation (\ref{depsi5}) we find
for axion 
\begin{equation}
h=h_0-\frac{4C_h}{(n-1)\beta _1R_m^{n-1}}\left( A_1^2z^{2\beta _1}+\beta
_2\right) ^{-1}  \label{hash5}
\end{equation}
with a new integration constant $h_0$. For the scale factor and synchronous
time coordinate $t_s$ of string frame from the relation (\ref{etostring}) we
have 
\begin{eqnarray}
\widetilde{R} &=&\widetilde{R}_0\left( z\frac{z^{\beta _1}+\beta _2z^{-\beta
_1}/A_1^2}{1+kz^2}\right) ^{1/(n-1)}  \label{erstring52} \\
t_s &=&t_{s0}+\frac{\pm 2}{n-1}\int_0^z\frac{\widetilde{R}\left( z\right) dz%
}{1+kz^2}.  \nonumber
\end{eqnarray}
As in the E - frame, here time coordinate $t_s$ varies in finite limits for
closed models and is unlimited for $k=0,-1$. For the models with zero scalar
fields $\psi _i$ the constant $\beta _1=\pm \sqrt{n}$ and from (\ref{erzet}%
), (\ref{e2phi5}) and (\ref{erstring52}) we obtain the solutions earlier
considered in \cite{copeland}.

In the limit $z\rightarrow 0$ all solutions tend to the solution with a flat
space, and in the E - frame 
\begin{equation}
R\sim \left| t_E-t_0\right| ^{1/n},\quad e^{2\varphi }\sim \left|
t_E-t_0\right| ^{\beta _3\frac{n-1}n},\quad t_E\rightarrow t_0
\label{asimp5}
\end{equation}
($\beta _3=-\left| \beta _1\right| $ for $\beta _2\neq 0$ and $\beta
_3=\beta _1$ for $\beta _2=0$ ). In the string frame one obtains 
\begin{equation}
\widetilde{R}\sim \left| t_s-t_{s0}\right| ^{\frac{1+\beta _3}{n+\beta _3}%
},\left| t_s-t_{s0}\right| =\left| t_E-t_0\right| ^{\frac{n+\beta _3}%
n},t_s\rightarrow t_{s0}  \label{asimp25}
\end{equation}
In this limit one has $R\rightarrow 0$ , while the limiting value of string
frame scale factor is zero or infinity for $\beta _3>-1$ and $\beta _3<-1$,
correspondingly. Note, that $\left| \beta _3\right| =\left| \beta _1\right|
\leq \sqrt{n}$. For the case $\beta _3=-1$ one has 
\begin{equation}
\widetilde{R}\sim const+\left( t_s-t_{s0}\right) ^2,\quad e^{2\varphi }\sim
\left| t_s-t_{s0}\right| ^{-1},\quad t_s\rightarrow t_{s0}  \label{treasymp}
\end{equation}
Despite the fact that in this case the string frame scale factor is finite
at $t_s=t_0$ the model is singular, because at that point the dilaton field
is divergent. Similar conclusion can be reached by starting directly with
the expression for scalar curvature, which in virtue of (\ref{hubble5}) can
be presented in the form 
\begin{equation}
R_{curv}=\frac{n(n-1)B}{R^{2n}}\sim \left( t_E-t_0\right) ^{-2}\sim \left|
t_s-t_{s0}\right| ^{\frac{-2n}{n-1}},\quad t_s\rightarrow t_{s0}
\label{Ricscal}
\end{equation}

We turn now to the other asymptotic limit $t_s\rightarrow \infty $ , which
is possible for open and flat models only. From the relations (\ref{e2phi5})
and(\ref{erstring52}) one can easily finds 
\begin{eqnarray}
\widetilde{R} &\sim &t_s^{\frac{1-\beta _3}{n-1}},\quad e^{2\varphi }\sim
t_s^{\frac{\beta _3(n-1)}{\beta _3-n}},\quad t_s\rightarrow \infty ,\quad k=0
\label{treasymp1} \\
\widetilde{R} &\sim &t_s,\quad e^{2\varphi }\sim \left( A_1+\beta
_2/A_1\right) /2+1/t_s^{n-1},\quad t_s\rightarrow \infty ,\quad k=-1 
\nonumber
\end{eqnarray}
Note that for $k=-1$ the limiting value $\varphi (\infty )>\varphi _{\min
}=\ln \beta _2/4$. For the relative contribution of moduli and antisymmetric
tensor field to the energy density one obtains 
\begin{equation}
\frac{n(n-1)B\varepsilon }{\stackrel{.}{\phi }^2R^2+\varepsilon }%
=\sum_iC_i^2+2C_h^2e^{-4\varphi }  \label{modener}
\end{equation}
and the evolution of the universe is moduli dominated at $t_s\rightarrow
t_{s0}$.The string frame scale factor $\stackrel{\sim }{R}$ and dilaton $%
e^\varphi $ (as functions of $t_s)$ in spatially flat, open and closed $n=3$
models are plotted in Fig.1 for $\left| \beta _1\right| <1$ (left hand-side)
and $\left| \beta _1\right| >1$ (right hand-side). The plots for the case $%
\left| \beta _1\right| =1$ can be found in \cite{behrndt} (see below).

For $n=3$ by introducing a new time coordinate $t$ in accordance with 
\begin{equation}
2kt^2=t_{+}^2+kt_{-}^2-\left( t_{+}^2-kt_{-}^2\right) \cos 2\sqrt{k}\eta
,\quad z=\sqrt{\frac{t^2-t_{-}^2}{t_{+}^2-kt^2}}  \label{newte}
\end{equation}
and choosing $R_m^2=\left( t_{+}^2-kt_{-}^2\right) /2$ for the line element
corresponding to the solution (\ref{er5}) we can write 
\begin{equation}
ds^2=\frac{t^2dt^2}{\sqrt{(t_{+}^2-kt^2)(t^2-t_{-}^2)}}-\sqrt{%
(t_{+}^2-kt^2)(t^2-t_{-}^2)}dl^2.  \label{interval5}
\end{equation}
For the particular values of integration constants 
\begin{equation}
\psi _{i1}=-1,\quad (\Rightarrow \beta _1=1)  \label{partval}
\end{equation}
from the expressions (\ref{psitree5}), (\ref{e2phi5}) and (\ref{hash5}) we
have 
\begin{eqnarray}
e^{2\varphi } &=&\frac{t^2\left( 1-kt_{-}^2/t_{+}^2\right) }{2\sqrt{%
(t_{+}^2-kt^2)(t^2-t_{-}^2)}},\quad e^{2\left( \psi _i-\psi _{i0}\right) }=%
\frac{t_{+}^2-kt^2}{t^2-t_{-}^2},  \label{partsol} \\
h &=&h_0\pm \frac{2t_{+}t_{-}}{t_{+}^2-kt_{-}^2}\left( \frac{t_{+}^2}{t^2}%
-k\right)  \nonumber
\end{eqnarray}
where the substitution $\beta _2=t_{-}^2/t_{+}^2$ is used. The expressions (%
\ref{interval5}), (\ref{partsol}) represent $4D$ tree level solutions of the
effective string theory in the E - frame. The corresponding solutions in the
string frame can be found by transformation (\ref{etostring}) with the
function $\widetilde{F}_R=e^{-2\varphi }$. They were found earlier and are
investigated in Ref. \cite{behrndt} by using the $5D$ solutions of \cite
{gibbons, horowitz}.

\section{ Conclusion}

In the present paper homogeneous anisotropic cosmological models are
investigated within the framework of the effective string gravity with
higher genus corrections and described by the action (\ref{actionst}). The
various conformal frames are considered, the most important of which are the
string , Einstein and Jordan frames. A class of nongravitational sources is
distinguished, for which the Jordan frame can be realized. The field
equations are derived in general conformal frame. It allows to obtain the
dynamics of model in various conformal frames by the corresponding choice of
the conformal factor. The equations of the cosmological model are the
simplest in the E - frame. In the absence of nongravitational sources the
solutions with Ricci - flat subspaces have the form (\ref{vacsol1}), (\ref
{dilontime1}) in the E - frame and (\ref{phierst}) in the string frame. They
generalize the multidimensional Kasner solution in GR. Further as an
additional source a set of modulus and Kalb - Ramond fields is considered.
In the case of zero potential terms this system is reduced to an effective
source with the extreme hard equation of state. For such a source the
equation of the dilaton field can be presented in the form (\ref{newform}).
The phase trajectories of the corresponding dynamical system in the
neighborhood of the solutions with constant dilaton are closed curves and
the nonmonotonic dilaton dependence of the coupling functions can not
stabilize the expectation value of the dilaton. This is a consequence of the
fact that for the source under consideration the corresponding friction term
in dilaton field equation vanishes. In the E- frame the corresponding
solution with a set of modulus and antisymmetric tensor fields as a
nongravitational source the scale factors coincide with those for vacuum
solution. For anisotropic models with Ricci - flat subspaces and for given
coupling functions in the Lagrangian (\ref{actionst}) the evolution of
dilaton, modulus and axion fields is determined by relations (\ref{solforphi}%
), (\ref{depsidete}). Section 5 is devoted to the isotropic models with a
space of arbitrary curvature and with a set of modulus fields and Kalb -
Ramond field as an additional source. The corresponding solutions have the
most simple form in terms of conformal time. The E - frame scale factor,
dilaton, modulus fields and axion are determined by relations (\ref{er5}), (%
\ref{phifunc5}) and (\ref{eps3}), and E - frame synchronous time coordinate
is connected with the conformal time by relation (\ref{te5}). Unlike the
flat and open models, lifetime of closed models is finite (see (\ref{liftime}%
)). The considered solutions are singular at some finite time moment. In a
vicinity of this point all solutions tend to the solution with Ricci - flat
subspaces. The tree level limit of received solutions is considered. The
corresponding formulas have the form (\ref{psitree5}), (\ref{e2phi5}), (\ref
{hash5}) in the E - frame and (\ref{erstring52}) in the string frame and are
generalizations of the earlier known solutions \cite{behrndt, caloper}.

For the solutions considered above the growth of the curvature is unbounded.
In the higher curvature regime string corrections modify the lowest - order
effective action (\ref{actionst}). In particular, higher derivative terms,
whose contribution is controlled by the string tension $\alpha ^{^{\prime }}$
, become important. The investigations have shown such corrections can
regularize the curvature singularity \cite{brust98}, \cite{anton} - \cite
{gaspven96}.

{\bf Acknowledgments}

The work is supported in part by Grant 96-855 of Ministry of Science and
Education of the Republic of Armenia. I would like to thank Prof. R.
Brustein for bringing the references \cite{madden1, brust98} to my attention.

\newpage \

\pagebreak \newpage

{\bf Figure Captions}

{\bf Fig.1}. Plots for the string frame scale factor $\stackrel{\sim }{R}$
(solid lines) and dilaton $e^\varphi $ (dashed)\nolinebreak as functions of
synchronous time coordinate $t_s$ in spatially flat (a,b), open (c,d) and
closed (e,f) $n=3$ models described by the solution (\ref{e2phi5}), (\ref
{erstring52}). On the left hand-side is $\left| \beta _1\right| <1$ and on
the right hand-side is $\left| \beta _1\right| >1$.

\end{document}